\documentclass[prx,aps,letterpaper,twocolumn,superscriptaddress,floatfix]{revtex4}
\usepackage{graphicx}
\usepackage{textcomp}
\usepackage{mathptmx}
\usepackage{amsmath}
\usepackage{xcolor}
\usepackage{soul}

\newcommand{\ket}[1]{\ensuremath{\left|#1\right\rangle}}
\definecolor{blue}{rgb}{0,0,1}
\definecolor{red}{rgb}{1,0,0}
\definecolor{green}{rgb}{0,1,0}

\begin{document}

\title{A two-fold quantum delayed-choice experiment in a superconducting
circuit}
\author{K.~Liu}
\author{Y.~Xu}
\author{W.~Wang}
\affiliation{Center for Quantum Information, Institute for Interdisciplinary Information
Sciences, Tsinghua University, Beijing 100084, China}
\author{Shi-Biao Zheng}
\email{t96034@fzu.edu.cn}
\affiliation{Department of Physics, Fuzhou University, Fuzhou 350116, China}
\author{Tanay Roy}
\author{Suman Kundu}
\author{Madhavi Chand}
\author{A.~Ranadive}
\author{R.~Vijay}
\affiliation{Department of Condensed Matter Physics and Materials Science, Tata Institute
of Fundamental Research, Homi Bhabha Road, Mumbai 400005, India}
\author{Y.~P.~Song}
\affiliation{Center for Quantum Information, Institute for Interdisciplinary Information
Sciences, Tsinghua University, Beijing 100084, China}
\author{L.-M. Duan}
\email{lmduanumich@gmail.com}
\affiliation{Center for Quantum Information, Institute for Interdisciplinary Information
Sciences, Tsinghua University, Beijing 100084, China}
\affiliation{Department of Physics, University of Michigan, Ann Arbor, Michigan 48109, USA}
\author{L.~Sun}
\email{luyansun@tsinghua.edu.cn}
\affiliation{Center for Quantum Information, Institute for Interdisciplinary Information
Sciences, Tsinghua University, Beijing 100084, China}

\begin{abstract}
We propose and experimentally demonstrate a two-fold quantum delayed-choice
experiment where wave or particle nature of a superconducting interfering
device can be post-selected twice after the interferometer. The
wave-particle complementarity is controlled by a quantum which-path detector
(WPD) in a superposition of its on and off states implemented through a
superconducting cavity. The WPD projected to its on state records which-path
information, which manifests the particle nature and destroys the
interference associated with wave nature of the system. In our experiment,
we can recover the interference signal through a quantum eraser even if the
WPD has selected out the particle nature in the first round of
delayed-choice detection, showing that a quantum WPD adds further
unprecedented controllability to test of wave-particle complementarity
through the peculiar quantum delayed-choice measurements. 
\end{abstract}

\maketitle

Wave-particle duality is among the most fundamental properties of quantum
mechanics. According to Bohr's principle of complementarity, a quantum
system possesses complementary and mutually exclusive properties that cannot
be observed at the same time~\cite{Bohr1984,Englert1996,Durr1998}. Whether a
quantum system behaves as a wave or a particle depends on the arrangement of
measurement apparatus, as demonstrated by the delayed-choice experiment with
a Mach-Zehnder (MZ) interferometer, in which the choice of inserting the
second beam splitter BS$_{2}$ or not is made after the photon has entered
the interferometer~\cite%
{Wheeler1984,Ma2014,Hellmuth1987,Lawson1996,Jacques2007,Manning2015}, as
shown in Fig.~1a. When BS$_{2}$ is inserted, no path information is
available so that the probability for detecting the photon at either outport
depends upon the relative phase $\varphi $ between the two paths, showing
the interference effect. In the absence of BS$_{2}$, detecting the photon at
each outport unambiguously reveals which route the photon has travelled and
no interference appears. The delayed-choice nature rules out the assumption
that the photon could know beforehand what type of experimental apparatus it
will be confronted with, and then behaves accordingly.

In a recent remarkable development, a quantum delayed-choice experiment was
proposed in Ref.~\cite{Ionicioiu2011}, where the action of BS$_{2}$ is
controlled by a quantum ancilla: When the ancilla is in the state $%
\left\vert 0\right\rangle _{a}$, BS$_{2}$ is removed; for the ancilla state $%
\left\vert 1\right\rangle _{a}$, BS$_{2}$ is inserted. The process, in terms
of quantum circuits~\cite{Nielsen}, is shown in Fig.~1b, where the Hadamard
gates H$_{1}$ and H$_{2}$ represent the corresponding beam splitters. When
the ancilla is in a superposition of $\left\vert 0\right\rangle _{a}$ and $%
\left\vert 1\right\rangle _{a}$, the wave-like and particle-like behaviors
of the test system can be observed at the same time; these two complementary
phenomena are encoded in the same output state and post-selected depending
on the measured value of the ancilla. This proposal has been successfully
demonstrated in experiments using photons \cite%
{Tang2012,Peruzzo2012,Kaiser2012} and other systems \cite%
{Roy2012,Auccaise2012,Zheng2015}.

\begin{figure*}[tbp]
\includegraphics{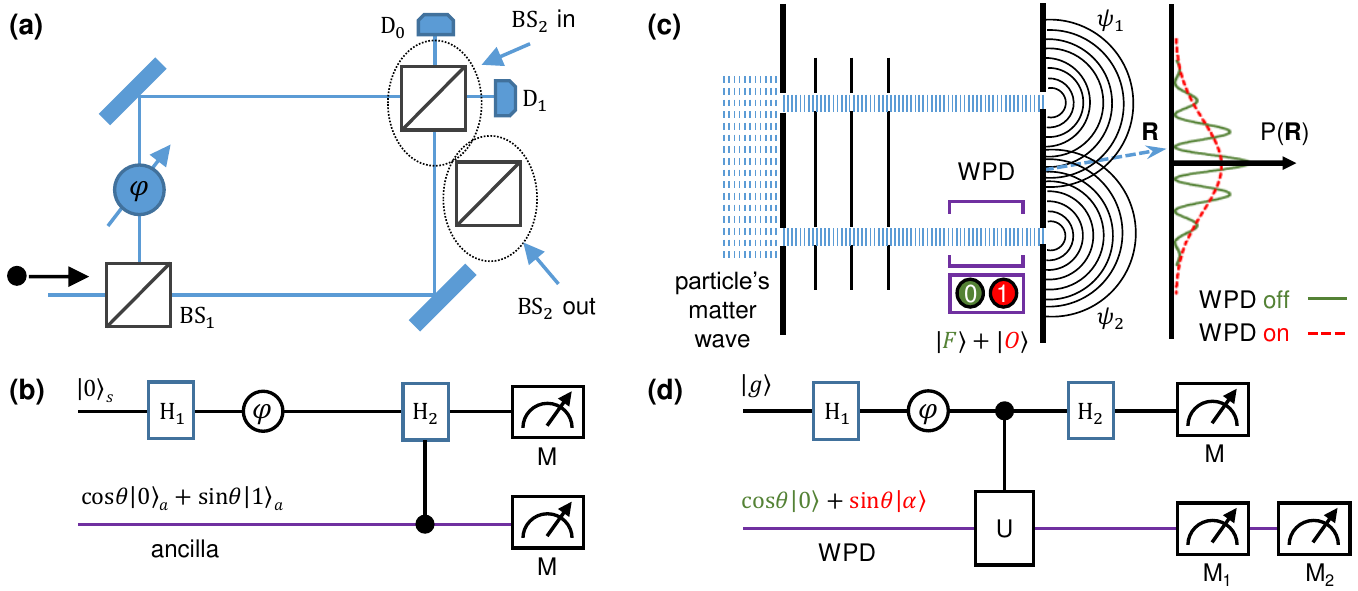}
\caption{\textbf{Schematic diagrams of delayed-choice experiments with a
two-path interferometer.} \textbf{(a)} Wheeler's delayed-choice thought
experiment. The decision of whether to insert BS$_2$ or not is delayed until
after the photon has entered the interferometer. \textbf{(b)} Quantum
delayed-choice experiment with a quantum beam splitter. Whether the second
beam splitter-like (Hadamard) operation, H$_2$, is applied or not depends
upon whether the ancilla is in the state $\left| 1\right\rangle _a$ or $%
\left| 0\right\rangle _a$. When the ancilla is in a superposition of these
states, the test system can simultaneously behave as a wave and a particle.
\textbf{(c)} Quantum delayed-choice experiment with a double-slit apparatus,
where a WPD is placed in front of slit 2. When the WPD is switched on, it
can store the which-path information of the incoming particle, destroying
the interference. If the WPD is switched off, no path information is
available. When the WPD is in a superposition of its on and off states, the
wave and particle behaviors are encoded in the same output state. \textbf{(d)%
} Quantum delayed-choice experiment using a Ramsey interferometer, where a
cavity acts as the WPD for a qubit's evolution in its quantum state space $%
\left\{ \left|g\right\rangle \text{, }\left| e\right\rangle \right\}$. When
the WPD contains a coherent field $\left| \protect\alpha \right\rangle $,
the conditional cavity $\protect\pi $-phase shift $U$ keeps track of the
which-path information in the field's phase. For the vacuum state $%
\left|0\right\rangle $, which-path information is not collected. The qubit's
particle and wave behaviors can be simultaneously observed by putting the
WPD in a superposition state. Two successive measurements on the WPD enable
implementation of a two-fold delayed-choice procedure.}
\end{figure*}

Here, we propose and experimentally demonstrate a two-fold quantum
delayed-choice experiment by introducing a WPD in a superposition of its on
and off states. As illustrated in Fig. 1c through a double-slit apparatus,
when the WPD is in its on state $\left\vert O\right\rangle $, it collects
the path information of the incoming quantum particle and labels its state
to $\left\vert O_{1}\right\rangle $ and $\left\vert O_{2}\right\rangle $
corresponding to the two different paths $1$ and $2$ of the particle with
wave functions $\psi _{1}(\mathbf{r})$ and $\psi _{2}(\mathbf{r})$,
respectively. When the WPD is in its off state $\left\vert F\right\rangle $,
it does not record any information of the incoming particle. If the WPD
starts at a superposition of its on and off states $\left( \left\vert
O\right\rangle +\left\vert F\right\rangle \right) /\sqrt{2}$, the combined
state of the WPD and the quantum particle becomes
\begin{equation}
\begin{aligned} \ensuremath{\left|\Phi (\mathbf{r})\right\rangle} =
\frac{1}{2}[(\ensuremath{\left|\psi
_{1}(\mathbf{r})\right\rangle}\ensuremath{\left|O_{1}\right\rangle}+%
\ensuremath{\left|\psi
_{2}(\mathbf{r})\right\rangle}\ensuremath{\left|O_{2}\right\rangle})\\
+(\ensuremath{\left|\psi
_{1}(\mathbf{r})\right\rangle}+\ensuremath{\left|\psi
_{2}(\mathbf{r})\right\rangle})\ensuremath{\left|F\right\rangle}].
\end{aligned}
\end{equation}%
If we measure the interference pattern of the quantum particle in the state $%
\left\vert \Phi (\mathbf{r})\right\rangle $, the interference fringes given
by the cross terms $\left\langle \psi _{1}(\mathbf{r})|\psi _{2}(\mathbf{r}%
)\right\rangle +\left\langle \psi _{2}(\mathbf{r})|\psi _{1}(\mathbf{r}%
)\right\rangle $ appear or disappear, depending on whether we post-project
the WPD's state to the vector $\left\vert F\right\rangle $ or the subspace $O
$ spanned by $\left\vert O_{1}\right\rangle $ and $\left\vert
O_{2}\right\rangle $. This gives a quantum delayed-choice experiment similar
to Refs.~%
\onlinecite{Ionicioiu2011,Roy2012,Auccaise2012,Tang2012,Peruzzo2012,Kaiser2012,Zheng2015}%
, but with the conceptual difference that this is enabled by a quantum WPD
in a classical interferometer instead of by a quantum interferometer. This
important difference leads to a distinction of the outcome: in our
experiment, we can later erase the which-path information of the quantum
particle~\cite{Scully1982,Scully1991,Gerry1996,Herzog1995,Kim2000,Ma2013} to
recover the interference fringes even after we have projected the WPD to the
\textquotedblleft on" subspace $O$. This is achieved by measuring the WPD in
the subspace $O$ along the basis $\left\vert \pm \right\rangle =\left(
\left\vert O_{1}\right\rangle \pm \left\vert O_{2}\right\rangle \right) /%
\sqrt{2}$, which fully restores the interference terms $Re\left[
\left\langle \psi _{1}(\mathbf{r})|\psi _{2}(\mathbf{r})\right\rangle \right]
$. In contrast, if we measure the WPD in the basis $\left\{ \left\vert
O_{1}\right\rangle ,\left\vert O_{2}\right\rangle \right\} $, the
interference fringe is completely lost. Therefore we can realize a two-fold
quantum delayed-choice experiment, where we have two chances to choose the
wave or particle behavior of the already detected quantum system, pushing
test of wave-particle complementarity to an unprecedented level of
controllability.

\begin{figure}[b]
\includegraphics{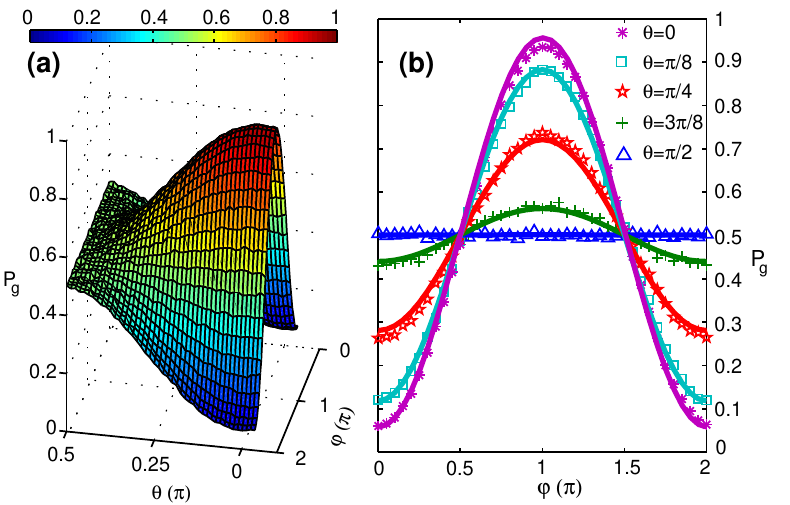}
\caption{\textbf{Ramsey interference pattern.} \textbf{(a)} Observation of
continuous morphing between the qubit's particle and wave behaviors. The
cavity, acting as the WPD, is initially prepared in the cat state of Eq.~3
with $\protect\alpha =2\protect\sqrt{2}$. The results clearly show that the
measured Ramsey interference pattern of the qubit, characterized by $P_g$ as
a function of $\protect\varphi $, depends upon the value of $\protect\theta $
that determines the state of the WPD. \textbf{(b)} Cuts in \textbf{(a)} for
different values of $\protect\theta $. Symbols are experimental data, which
are in good agreement with the numerical simulations (solid lines) that take
into account the experimental imperfections (except for measurement errors)
including the qubit and cavity decoherence, the cross-Kerr interaction
between the qubit and the cavity, and the self-Kerr interaction of the
cavity. The standard deviation for each measured value is less than $0.01$
and not shown in the figure.}
\end{figure}

We implement the two-fold quantum delayed-choice experiment with a
superconducting Ramsey interferometer. The circuit diagram is shown in
Fig.~1d, which is similar to that of a MZ interferometer, with the Hadamard
gates H$_{1}$ and H$_{2}$ representing the beam splitters for the quantum
state of a qubit, whose basis vectors $\left\vert g\right\rangle $ and $%
\left\vert e\right\rangle $ correspond to the two paths in the MZ
interferometer. Here both H$_{1}$ and H$_{2}$ are implemented by a classical
microwave pulse, corresponding to $\pi /2$ rotations on the Bloch sphere.
The qubit, initially in the state $\left\vert g\right\rangle $, is
transformed to the superposition state $(\left\vert g\right\rangle
+\left\vert e\right\rangle )/\sqrt{2}$ by H$_{1}$. Then a relative phase
shift between $\left\vert g\right\rangle $ and $\left\vert e\right\rangle $
is introduced to get the state $(\left\vert g\right\rangle +e^{i\varphi
}\left\vert e\right\rangle )/\sqrt{2}$, mimicking the relative phase $%
\varphi $ between the two interferometer arms. The WPD is represented by a
cavity in the dispersive region, with its coupling to the qubit described by
the Hamiltonian
\begin{equation}
H_{e}=\hbar \chi _{qs}a^{\dagger }a\otimes \left\vert e\right\rangle
\left\langle e\right\vert ,
\end{equation}%
where $\chi _{qs}$ denotes the coupling rate, and $a^{\dagger }$ and $a$ are
the creation and the annihilation operators for the cavity mode. At
interaction time $\tau =\pi /\chi _{qs}$, this coupling leads to a
qubit-state-dependent $\pi $-phase shift: $U=e^{i\pi a^{\dagger }a\otimes
\left\vert e\right\rangle \left\langle e\right\vert }$~\cite%
{Deleglise,Vlastakis,SunNature}. When the cavity is in the vacuum state $%
\left\vert 0\right\rangle $ (corresponding to the off state $\left\vert
F\right\rangle $ of the WPD), the qubit state is not affected by the cavity
coupling operator $U$. After the Hadamard gate H$_{2}$, the probability for
recording the qubit in the state $\left\vert g\right\rangle $ is given by $%
P_{g}=(1-\cos \varphi )/2$, which shows a perfect interference fringe with
respect to the relative phase $\varphi $. However, if the cavity starts in a
coherent state $\left\vert \alpha \right\rangle $ (corresponding to the on
state $\left\vert O\right\rangle $ of the WPD), the coupling operator $U$
evolves the qubit-cavity system to the entangled state $(\left\vert
g\right\rangle \left\vert \alpha \right\rangle +e^{i\varphi }\left\vert
e\right\rangle \left\vert -\alpha \right\rangle )/\sqrt{2}$. The probability
for getting the qubit in the state $\left\vert g\right\rangle $ after H$_{2}$
is given by $P_{g}=(1-e^{-2\left\vert \alpha \right\vert ^{2}}\cos \varphi
)/2$. When the overlap of the two labelling states $\left\vert \alpha
\right\rangle $ and $\left\vert -\alpha \right\rangle $ is negligible, i.e.,
$e^{-2\left\vert \alpha \right\vert ^{2}}\ll 1$, the interference fringe
disappears. In our experiment, the cavity is prepared in a superposition cat
state
\begin{equation}
\left\vert \phi \right\rangle =\cos \theta \left\vert 0\right\rangle +\sin
\theta \left\vert \alpha \right\rangle ,
\end{equation}%
and the probability for recording the qubit in state $\left\vert
g\right\rangle $ is given by
\begin{equation}
P_{g}\approx \lbrack \cos ^{2}\theta (1-\cos \varphi )+\sin ^{2}\theta ]/2,
\end{equation}%
where the two terms of $P_{g}$ show respectively the wave behavior
(interference) and the particle behavior (no interference) of the
superconducting qubit.

The experimental implementation is based on a three-dimensional circuit
quantum electrodynamics (circuit QED) architecture~\cite{Paik} (see
Supplementary Material for details). The experiment starts with preparing
the cavity in a coherent superposition of $\left\vert 0\right\rangle $ and $%
\left\vert \alpha \right\rangle $ as in Eq.~3. The experimental pulse
sequence (in Fig.~\ref{fig:figS3}) and detailed description are presented in the
Supplementary Material. We experimentally generate such states with $\alpha
=2\sqrt{2}$. For $\theta =\pi /4$, the fidelity $F=\left\langle \phi
\right\vert \rho _{p}\left\vert \phi \right\rangle =0.93$, where $\rho _{p}$
is the density operator of the produced cavity state (measured Wigner
function is shown in Fig.~\ref{fig:figS4} in the Supplementary Material). After creating
the cat state, we sandwich the conditional $\pi $-phase shift gate $U$
between two Hadamard gates H$_{1}$ and H$_{2}$ for the Ramsey interference
measurement. In our experiment the tunable phase shift $\varphi $ is
incorporated into the first Hadamard gate H$_{1}$ by adjusting the phase of
the corresponding microwave pulse. In Figs. 2a and 2b, we present the
measured probability $P_{g}$\ as a function of $\varphi $\ for different
values of $\theta $. As expected, the qubit exhibits a continuous transition
between wave-like and particle-like behavior by varying $\theta $ that
determines the initial state of the WPD.

\begin{figure}[tbp]
\includegraphics{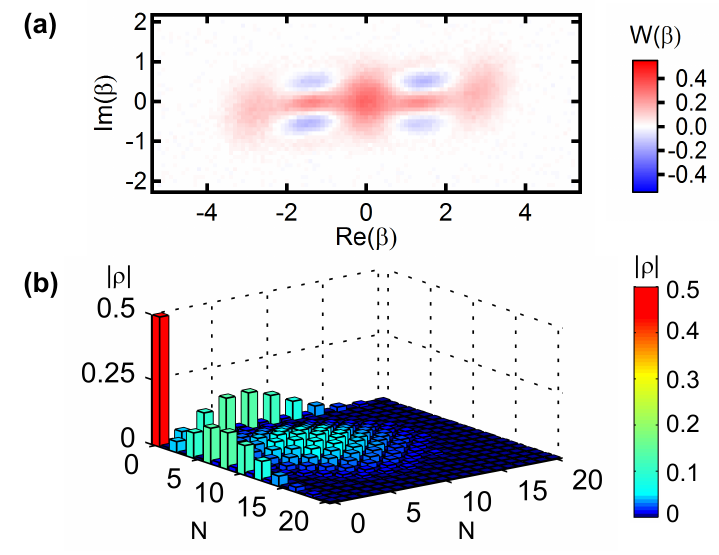}
\caption{\textbf{Reconstruction of the cavity states after the second Ramsey
pulse without selection on the qubit state.} \textbf{(a)} The measured
Wigner function and \textbf{(b)} density matrix for $\protect\alpha =2%
\protect\sqrt{2}$, $\protect\theta =\protect\pi /4$ and $\protect\varphi =%
\protect\pi /2$. To clearly illustrate the quantum coherence between $\left|
0\right\rangle $ and $\left| \protect\alpha \right\rangle $ ($\left| -%
\protect\alpha \right\rangle $), we here only present the moduli of the
density matrix elements in the Hilbert space, which are obtained from the
corresponding measured Wigner function. The overlap (fidelity) between the
measured density matrix $\protect\rho _m$ and ideal result $\protect\rho _i$%
, defined as $F=[Tr(\protect\sqrt{\protect\sqrt{\protect\rho _i}\protect\rho %
_m\protect\sqrt{\protect\rho _i}})]^2$, is about $0.80$; the infidelity is
mainly due to the finite bandwidth of the $\protect\pi /2$ qubit pulses.}
\end{figure}

\begin{figure}[tbp]
\includegraphics{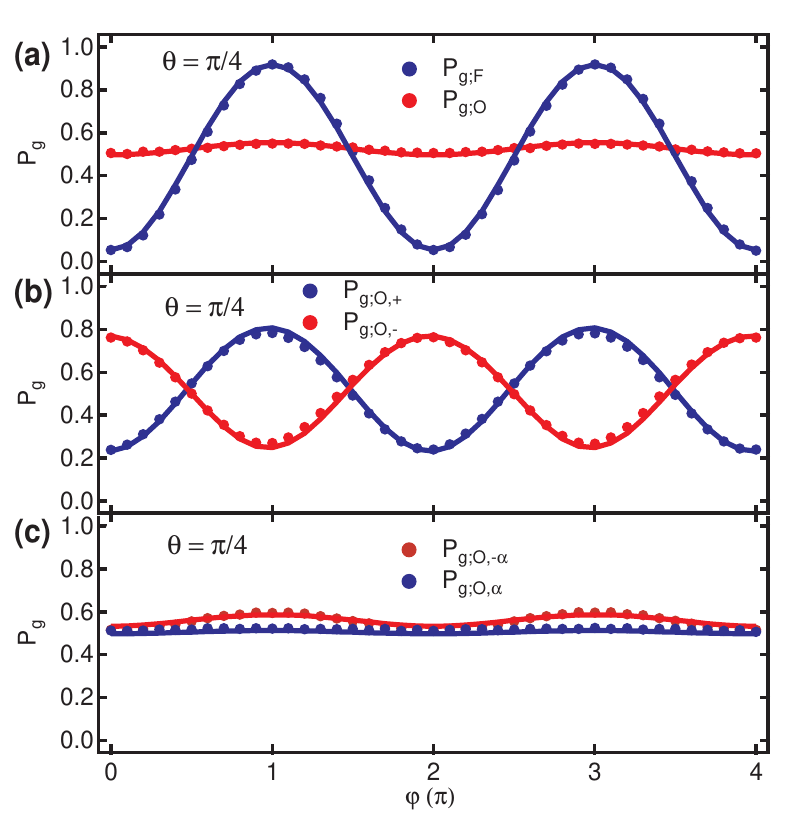}
\caption{\textbf{Conditional Ramsey interference signals by post-selection
on the WPD's state.} \textbf{(a)} The probabilities $P_{g;O}$ and $P_{g;F}$,
as functions of $\protect\varphi $, for detecting the qubit in the state $%
\left| g\right\rangle $ in the Ramsey interference experiment conditional
upon the post-selection of the cavity's states $\left| \pm \protect\alpha %
\right\rangle $ and $\left| 0\right\rangle $, respectively. \textbf{(b)} The
probabilities $P_{g;O,+}$ and $P_{g;O,-}$, versus $\protect\varphi $, for
measuring the qubit in the state $\left| g\right\rangle $, conditional upon
the measurement of even and odd parities of the WPD following the detection
of the on state, respectively. \textbf{(c)} The probabilities $P_{g;O,%
\protect\alpha }$ and $P_{g;O,-\protect\alpha }$, for detecting the qubit in
the state $\left| g\right\rangle $, conditional upon post-selecting the
WPD's on state and subsequently detecting the components $\left| \protect%
\alpha \right\rangle $ and $\left| -\protect\alpha \right\rangle $,
respectively. The parameters are $\protect\alpha =2\protect\sqrt{2}$ and $%
\protect\theta =\protect\pi /4$. Dots represent experimental data with the
standard deviation less than 0.01 and not shown, in excellent agreement with
the numerical simulations (solid lines) based on the measured device
parameters.}
\end{figure}

The essence of quantum delayed-choice experiments is that the quantum
coherence of the measuring device excludes the possibility of the qubit
knowing the measurement choice in advance. To verify the presence of the
quantum coherence between the on and off states of the WPD, we perform the
Wigner tomography after the Ramsey interference experiment~\cite%
{Deleglise,Vlastakis,SunNature,Lutterbach1997}. The Wigner function of a
quantum harmonic oscillator, the quasiprobability distribution in phase
space, contains all the information of the associated quantum state~\cite%
{Schleich2001}. In Fig.~3 we present the reconstructed state of the cavity
without post-selection on the qubit state (data conditional upon the qubit
state $\left|g\right\rangle $ and $\left|e\right\rangle $ are presented in
Fig.~\ref{fig:figS5} in the Supplementary Material) for $\theta =\pi /4$ and $\varphi
=\pi /2$. As expected, the measured Wigner function (Fig.~3a) exhibits
interference fringes, with alternate positive and negative values between $0$
and $\alpha$ ($-\alpha $) in phase space, which are the signature of quantum
coherence between $\left\vert 0\right\rangle $ and $\left\vert \alpha
\right\rangle $ ($\left\vert -\alpha \right\rangle $). To further
characterize this coherence, we display the density matrix (Fig.~3b) in the
Hilbert space obtained from the measured Wigner function, where each term
represents the modulus of the corresponding matrix element (without
including the phase). The quantum coherence between $\left\vert
0\right\rangle $ and $\left\vert \alpha \right\rangle $ ($\left\vert -\alpha
\right\rangle $) is manifested in the off-diagonal elements $\left\vert \rho
_{n,0}\right\vert $ and $\left\vert \rho _{0,n}\right\vert $, which are
responsible for the observed interference fringes in the Wigner function~%
\cite{Deleglise}. These results unambiguously demonstrate that the qubit's
behaviors with and without interference are observed with the same setup,
where the WPD is in a quantum superposition of its on and off states.

To post-select the qubit's behavior, we examine whether the cavity is filled
with a coherent field or is empty by using a conditional qubit rotation. Due
to the dispersive qubit-cavity coupling, the qubit's transition frequency
depends on the photon number of the cavity field, so that we can drive the
qubit's transition conditional on the cavity being in a specific Fock state.
After the Ramsey interference experiment we apply a $\pi $ pulse to the
qubit conditional on the cavity's vacuum state: $R_{\pi ,0}^{Y}=\exp \left[
\frac{\pi }{2}\left\vert 0\right\rangle \left\langle 0\right\vert \otimes
\left( \left\vert e\right\rangle \left\langle g\right\vert -\left\vert
g\right\rangle \left\langle e\right\vert \right) \right] $. Since both $%
\left\vert \alpha \right\rangle $ and $\left\vert -\alpha \right\rangle $
are approximately orthogonal to $\left\vert 0\right\rangle $, the qubit
undergoes no transition for these two coherent state components. It should
be noted that the coherent state $\left\vert \alpha \right\rangle $ is not
distinguished from $\left\vert -\alpha \right\rangle $ after this procedure,
that is, the which-path information is not read out. In Fig.~4a we present
the measured conditional probabilities $P_{g;O}$\ and $P_{g;F}$ as functions
of $\varphi $, which are defined as the probabilities for detecting the
qubit in $\left\vert g\right\rangle $ in the Ramsey interference experiment
conditional on the measurement of the WPD's on state $\left\vert \pm \alpha
\right\rangle $ and off state $\left\vert 0\right\rangle $, respectively.
Here the on and off states have equal weighting, i.e., $\theta =\pi /4$. $%
P_{g;F}$\ clearly manifests the qubit's interference behavior with a
visibility of $0.89$, while $P_{g;O}$ almost shows no dependence on $\varphi
$, as expected. Due to the imperfection of the conditional $\pi $ pulse
(Fig.~\ref{fig:figS6}g in the Supplementary Material), there is a small probability that $%
\left\vert 0\right\rangle $ is taken for $\left\vert \pm \alpha
\right\rangle $; this accounts for the residual interference effects (with a
fringe contrast of $0.04$) in $P_{g;O}$. These results clearly demonstrate
that the qubit's behavior depends upon the WPD's state; the quantum
coherence of the on and off states, as shown in Fig.~3, excludes any model
in which the choice corresponds to a classical variable that has been known
in advance.

Unlike the previous quantum delayed-choice experiments~\cite%
{Ionicioiu2011,Roy2012,Auccaise2012,Tang2012,Peruzzo2012,Kaiser2012,Zheng2015}%
, here the which-path information associated with the qubit's particle-like
behavior is not read out during the observation of the interference pattern;
instead, it is stored in the field's phase. This path distinguishability can
be erased: When we measure the cavity's parity, instead of distinguishing
between $\left| 0\right\rangle $ and $\left| \pm \alpha \right\rangle $, and
correlate the outcomes with the qubit's data, two complementary interference
patterns appear (Fig.~\ref{fig:figS7} in the Supplementary Material). More intriguingly,
the quantum erasure~\cite%
{Scully1991,Gerry1996,Scully1982,Herzog1995,Kim2000,Ma2013} can be realized
even after the particle-like behavior (the red line of Fig.~4a) has been
post-selected, resulting in the restoration of the fringes. In Fig.~4b we
plot the measured conditional probabilities $P_{g;O,+}$\ and $P_{g;O,-}$ as
functions of $\varphi $, defined as the probabilities for measuring the
qubit in $\left| g\right\rangle $\ after H$_2 $, conditional on the
detection of even and odd parities of the WPD following the post-selection
of its on state, respectively. The fringe contrasts associated with $%
P_{g;O,+}$\ and $P_{g;O,-}$ are 0.53 and 0.48, respectively. The
imperfection of these conditional interference patterns is mainly due to the
infidelity of the Hadamard gates used for qubit Ramsey interference and WPD
parity discrimination, which is predominantly caused by cavity photons.

On the other hand, if we choose to read out the which-path information by
distinguishing between $\left\vert \alpha \right\rangle $ and $%
\left\vert-\alpha \right\rangle $, the qubit shows no interference, as shown
in Fig.~4c, where $P_{g;O,\alpha }$ and $P_{g;O,-\alpha }$ denote the
probabilities to detect the state $\left\vert g\right\rangle $\ after H$_{2}$%
, conditional on the post-selection of the WPD's on state and the subsequent
measurement of $\left\vert \alpha \right\rangle $ and $\left\vert -\alpha
\right\rangle $, respectively. These two states can be distinguished by
successively performing the cavity displacement $D(-\alpha )$, the
conditional qubit $\pi $ rotation $R_{\pi ,0}^{Y}$, and the qubit state
measurement. We note that thus obtained $\left\vert -\alpha \right\rangle $
state is mixed up with the residual vacuum, resulting in a slight
oscillation in $P_{g;O,-\alpha }$ with a contrast of $0.07$. This fringe
contrast can be further reduced by performing additional operations $%
D(\alpha ) $ and $R_{\pi ,0}^{Y}$ and then measuring the qubit state to
remove the residual vacuum. The deviation of $P_{g;O,-\alpha }$ from $1/2$
is mainly due to the qubit energy decay (see Supplementary Material).

The delayed-choice quantum eraser embedded in the quantum delayed-choice
experiment is only enabled by employing the quantum properties of the WPD,
significantly extending the concept of delayed-choice experiment. The
two-fold delayed-choice procedure provides a clear demonstration that the
behavior with or without interference is not a realistic property of the
test system: It depends not only on the delayed choice of the WPD's state,
but also on how we later measure the WPD and correlate the outcomes with the
data of the test system.

In summary, we have proposed a two-fold quantum delayed-choice experiment
which is enabled by using a WPD prepared in a superposition of its on and
off states. We implemented the experiment in circuit QED, observing both
behaviors with and without interference for a superconducting qubit in the
same experiment by using a cavity in a cat state as the WPD. We confirmed
the existence of quantum coherence between the on and the off states of the
WPD, excluding interpretations of the results based on classical models. The
quantum properties of the WPD allows erasure of the which-path information
associated with the post-selected particle-like behavior, implementing a
two-fold delayed-choice procedure and illustrating the wave-particle
complementarity in an unprecedented manner.

\begin{acknowledgments}
We thank M.~H.~Devoret for helpful discussions. This work was supported by
the Ministry of Science and the Ministry of Education of China through its
grant to Tsinghua University, the National Natural Science Foundation of
China under Grant Nos. 11374054 and 11474177, the Major State Basic Research
Development Program of China under Grant No. 2012CB921601, and the 1000
Youth Fellowship program in China.
\end{acknowledgments}


\appendix
\section{Experiment setup and device parameters}
\begin{figure}[b]
\includegraphics{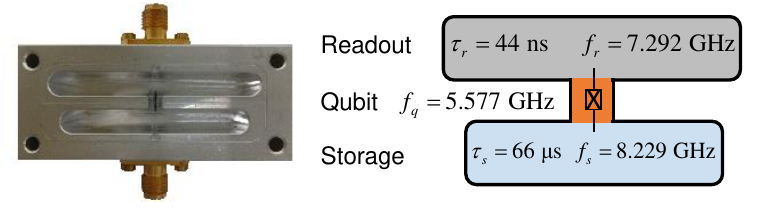}
\caption{\textbf{Experimental device.} Optical image and schematic of the device, where a transmon qubit is located in a trench and coupled to two 3D Al cavities.}
\label{fig:figS1}
\end{figure}

Our experiment is implemented with a three-dimensional circuit QED architecture~\cite{Paik} as shown in Fig.~\ref{fig:figS1}, where a single transmon qubit in a waveguide trench is dispersively coupled to two 3D cavities~\cite{Vlastakis,SunNature}, one cavity serving as the storage cavity and the other used to read out the qubit's state. The transmon qubit is fabricated on a $c$-plane sapphire (Al$_2$O$_3$) substrate with a double-angle evaporation of aluminum after a single electron-beam lithography step. The qubit has a transition frequency $\omega _q/2\pi =5.577$ GHz with an anharmonicity $\alpha _q/2\pi=(\omega _{ge}-\omega _{ef})/2\pi =246$ MHz, an energy relaxation time $T_1=9.5~\mu $s, a Ramsey time $T_2^{*}=7.5~\mu $s, and a pure dephasing time $T_\phi =12.4~\mu $s. The sample is placed in a cryogen-free dilution refrigerator at a base temperature of about 10~mK. Even at the lowest base temperature, the qubit is measured to have a probability about 8.5\% of being populated in the excited state $\ket{e}$ in the steady state. The exact source for this excitation is unknown; it may be caused by stray infrared photons or other background noise leaking into the cavity. This excitation can be removed through an initialization measurement of the qubit state by post-selecting the projections of the system onto the ground state $\left| g\right\rangle$~\cite{Riste2012} (also see below). Both the storage and readout cavities are made of aluminum alloy 6061 with a frequency of 8.229 GHz and 7.292 GHz, respectively. The photon lifetimes in the storage and readout cavities are $\tau _s=66~\mu $s and $\tau _r=44$ ns, respectively. The dispersive couplings between the qubit and the storage (readout) cavity is $\chi _{qs}/2\pi =-1.64$~MHz ($\chi _{qr}/2\pi =-4.71$~MHz). For simplicity, we will refer to the storage cavity as ``the cavity'' henceforth. This cavity acts as the which-path detector (WPD) for the qubit in the Hilbert space. 

\begin{figure*}
\includegraphics{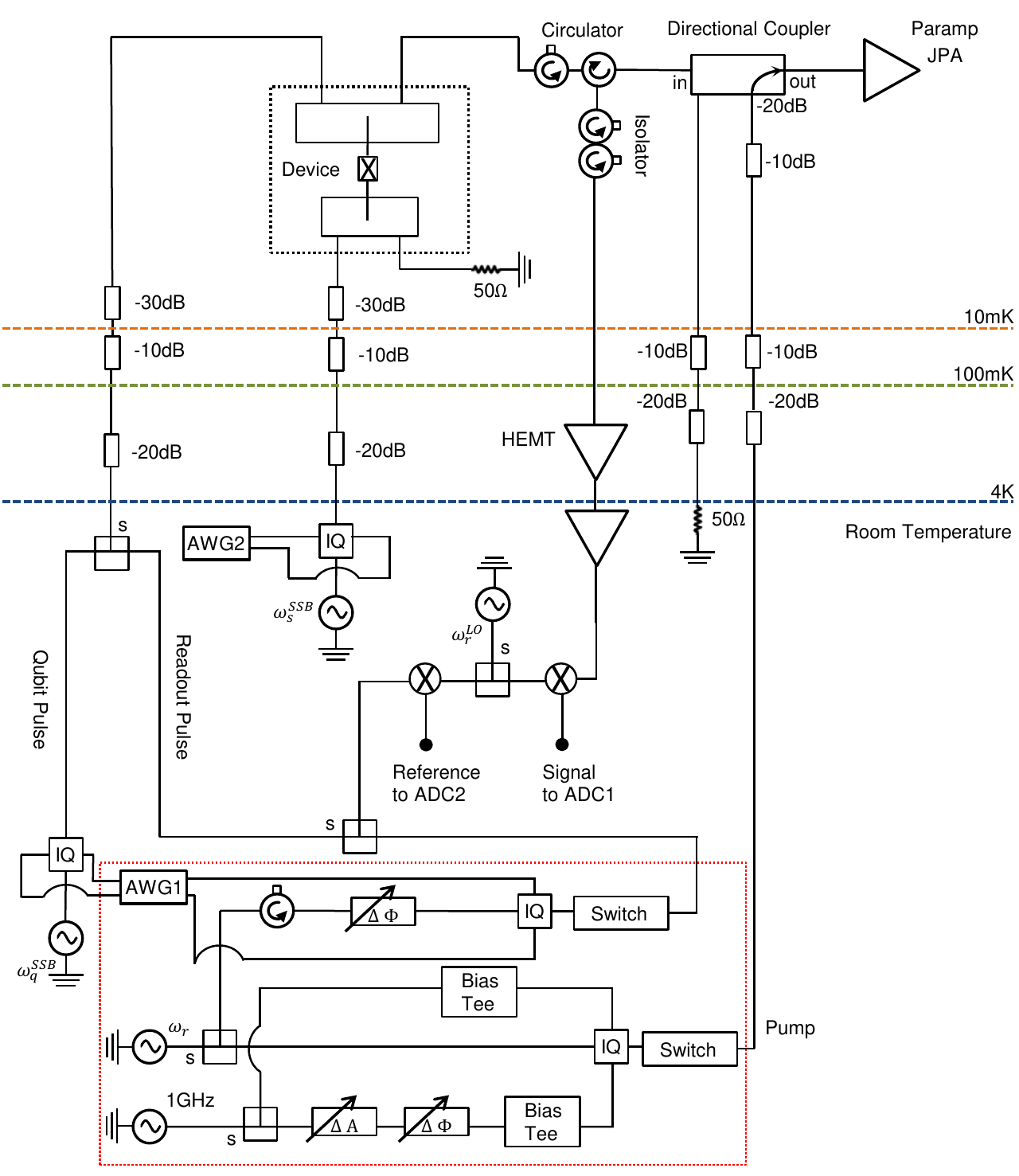}
\caption{\textbf{Schematic of the measurement setup.} A JPA operating in a double-pumped mode is used to read out the qubit state with a high fidelity and in a QND way. The part within the dashed red rectangle shows the biasing of the JPA.}
\label{fig:figS2}
\end{figure*}

The schematic of the measurement setup is shown in Fig.~\ref{fig:figS2}. We apply a Josephson parametric amplifier (JPA)~\cite{Hatridge,Roy} operating in a double-pumped mode~\cite{Kamal,Murch} (red enclosed part in Fig.~\ref{fig:figS2} shows the biasing circuit) as the first stage of amplification between the readout cavity at base and the high-electron-mobility-transistor amplifier at 4K. To minimize pump leakage into the readout cavity for a longer $T_\phi$ time, we typically operate the JPA in a pulsed mode. This JPA allows for a high-fidelity and quantum non-demolition (QND) single-shot readout of the qubit state (see the section on ``Readout Properties'').

\begin{figure*}
\includegraphics{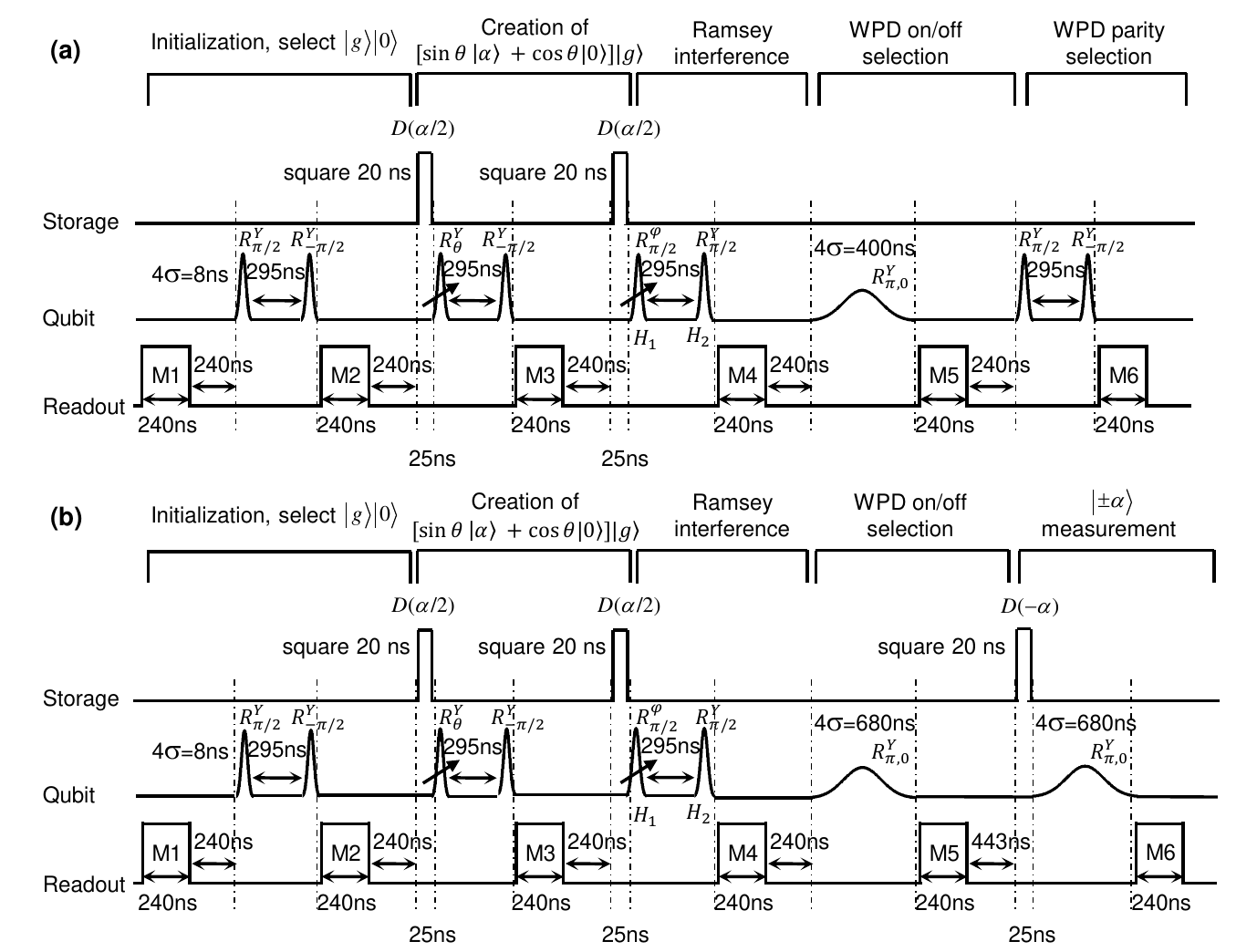}
\caption{\textbf{Schematic of the pulse sequences in the experiments.} The whole experiment can be divided into four main parts: 1) initialization of the system to $\ket{g}\ket{0}$ by post-selecting results of the qubit measurement and the cavity parity measurement; 2) creation of an arbitrary superposition state of the WPD, ${\sin\theta}\ket{\alpha}+{\cos\theta}\ket{0}$; 3) Ramsey interference measurement of the qubit with variable phase $\varphi$ in the first $\pi/2$ pulse $R^{\varphi}_{\pi/2}$; 4) measurements of the WPD's state: on/off state post-selection of the WPD (gives $P_{g;O}$ and $P_{g;F}$, Fig.~4a in the main text), and, under the on state, \textbf{(a)} the parity measurement of the WPD (gives $P_{g;O,+}$ and $P_{g;O,-}$, Fig.~4b in the main text); \textbf{(b)} the discrimination of $\ket{\alpha}$ and $\ket{-\alpha}$ states (gives $P_{g;O,\alpha}$ and $P_{g;O,-\alpha}$, Fig.~4c in the main text).}
\label{fig:figS3}
\end{figure*}

\begin{figure*}
\includegraphics{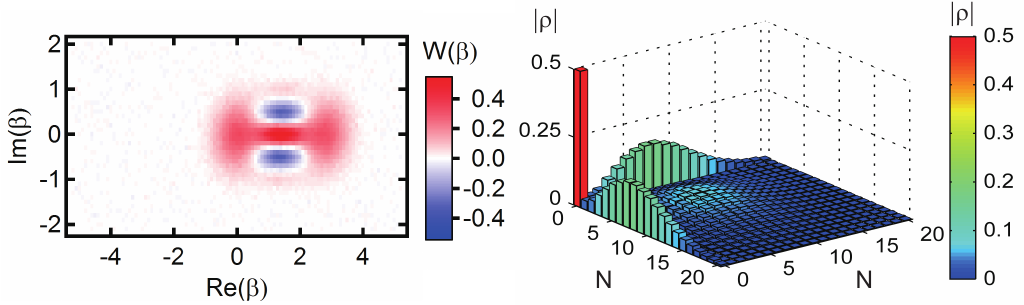} 
\caption{\textbf{Tomography of initial cavity state.} Measured Wigner function (left) and density matrix (right) of the cavity initially prepared in the cat state of $\ket{\phi}=\mathnormal{\cos\theta}\ket{0}+\mathnormal{\sin\theta}\ket{\alpha}$ (Eq.~3 in the main text) with $\alpha=2\sqrt{2}$ and $\theta=\pi/4$. The fidelity $F=\left\langle \phi \right| \rho _p\left| \phi \right\rangle =0.93$, where $\rho _p$ is the density operator of the produced cavity state.}
\label{fig:figS4}
\end{figure*}

\begin{figure*}
\includegraphics{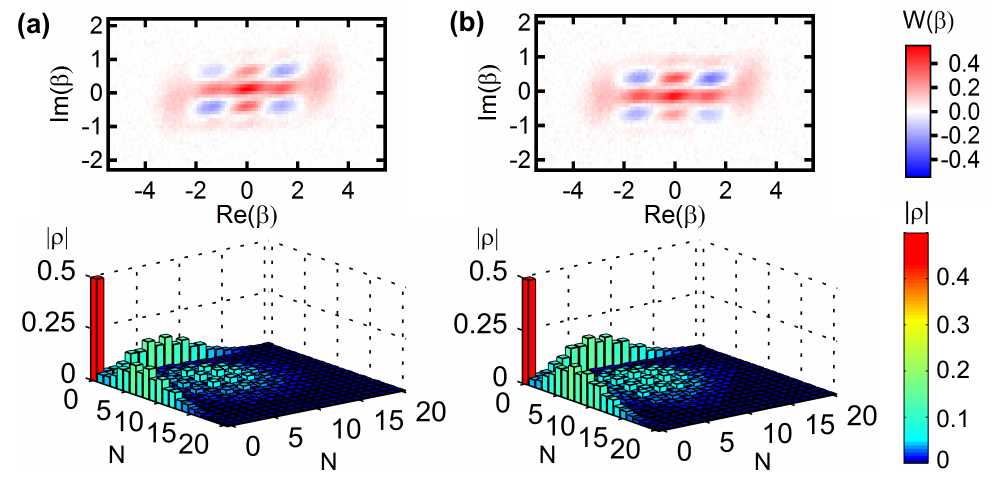} 
\caption{\textbf{Reconstruction of the cavity states after the second Ramsey
pulse.} \textbf{(a)} Conditional upon the qubit state $\left| g\right\rangle
$. \textbf{(b)} Conditional upon the $\left| e\right\rangle $ state. The measured Wigner functions and
density matrices are shown in the upper and lower rows, respectively, all
for $\protect\alpha =2\protect\sqrt{2}$, $\protect\theta =\protect\pi /4$
and $\protect\varphi =\protect\pi /2$. To clearly illustrate the quantum
coherence between $\left| 0\right\rangle $ and $\left| \protect\alpha %
\right\rangle $ ($\left| -\protect\alpha \right\rangle $), we here only
present the moduli of the density matrix elements in the Hilbert space,
which are obtained from the corresponding measured Wigner functions. The
overlap (fidelity) between the measured density matrix $\protect\rho _m$ and
ideal result $\protect\rho _i$, defined as $F=[Tr(\protect\sqrt{\protect%
\sqrt{\protect\rho _i}\protect\rho _m\protect\sqrt{\protect\rho _i}})]^2$,
is about $0.80$ for each case; the infidelities are mainly due to the finite
bandwidth of the $\protect\pi /2$ qubit pulses.}
\label{fig:figS5}
\end{figure*}

\section{Pulse sequence and parameters}
\begin{figure*}
\includegraphics{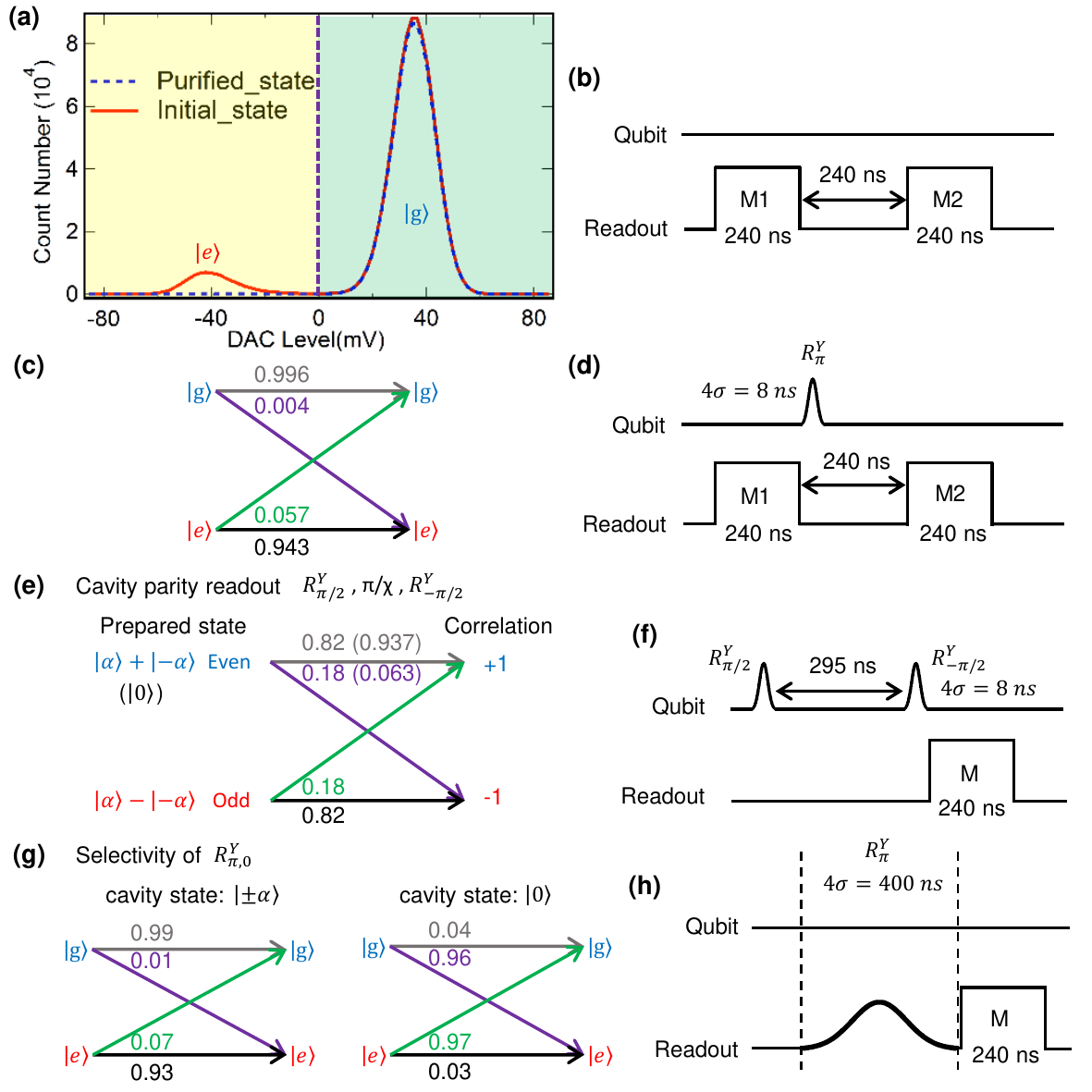}
\caption{\textbf{Readout properties.} \textbf{(a)} Bimodal and well-separated histogram of the qubit readout. A threshold $V_{th}=0$ is chosen to digitize the readout signal. The solid line corresponds to the case without the qubit purification measurement, showing the qubit has an 8.5\% probability of being populated in $\ket{e}$ state, while the dashed line represents the result after the qubit has been purified. \textbf{(b)} The measurement pulse sequence used in \textbf{a}. \textbf{(c)} Basic qubit readout matrix with the cavity left in vacuum. The readout errors are mainly due to the $T_1$ process. \textbf{(d)} The measurement pulse sequence used in \textbf{c}. \textbf{(e)} Readout property of the cavity parity state. The parity readout infidelity for the WPD's on state ($\alpha=2\sqrt{2}$) is mainly due to the loss of effectiveness of the $\pi/2$ pulses and the qubit decoherence. \textbf{(f)} The pulse sequence for the parity measurement. \textbf{(g)} The selectivity property of the Gaussian $\pi$ pulse $R_{\pi,0}^Y$ with $\sigma=100$~ns. The difference between the readout fidelities of the cavity state $\ket{\pm \alpha=\pm 2\sqrt2}$ associated with the qubit states $\ket{g}$ and $\ket{e}$ is due to the decay of $\ket{e}$ during the long duration of this selective $\pi$ pulse. \textbf{(h)} The pulse sequence used in \textbf{g}.}
\label{fig:figS6}
\end{figure*}

The pulse sequence for the delayed-choice experiment is shown in Fig.~\ref{fig:figS3}. The whole process can be divided into four main parts: initialization of the system, creation of the superposition state of the WPD, Ramsey interference measurement of the qubit, and different measurements of the WPD's state. It is worth mentioning that except for the first two stochastic measurement parts, all the following measurement results are recorded and included in the analysis. 

To get a high signal-to-noise ratio, the readout pulse is optimized to contain a few photons with a width of 240~ns. The first measurements in Figs.~\ref{fig:figS3}a and \ref{fig:figS3}b are to purify the qubit state to $\ket{g}$ by a post-selection. The pulses used to perform phase-space displacement for the cavity mode is chosen to be square-shaped with a width of 20~ns for a broad frequency span. The displacement amplitude is calibrated by a global Poisson distribution fitting of the probability of the first eight Fock states as a function of the arbitrary waveform generator DAC output~\cite{SunNature}. This calibration also results in a background photon population $n_{th}\sim 0.01$ in the cavity. In order to eliminate this thermal excitation, a cavity parity measurement is performed at the beginning of each experiment to purify the cavity state to $\ket{0}$ (M2 in Figs.~\ref{fig:figS3}a and \ref{fig:figS3}b).

The cavity parity measurement is achieved by sandwiching a conditional cavity $\pi $-phase shift between two qubit $\pi/2$ rotations along the $Y$-axis on the Bloch sphere but with opposite rotation axes, denoted as $R_{\pm \pi/2}^Y$~\cite{Bertet,Deleglise,Vlastakis,SunNature}. To achieve a high-fidelity parity measurement, the $R_{\pm \pi/2}^Y$ pulses used to perform these qubit rotations should equally cover as many number splitting peaks as possible without significantly exciting the second and higher excited state. As a good compromise, we use Gaussian envelope pulses truncated to $4\sigma =8~ns$ ($\sigma _f=80$~MHz). We also apply the so-called ``derivative removal by adiabatic gate" (DRAG) technique~\cite{Motzoi} to eliminate the possible qubit leakage to higher levels.

To distinguish a vacuum state $\ket{0}$ and a coherent state $\ket{\alpha}$ ($\alpha=2\sqrt{2}$ in our experiment), we apply a conditional qubit $\pi$ pulse $R_{\pi,0}^Y$ which is effective only for a vacuum state. This pulse has a Gaussian envelope with $\sigma \geq 100$~ns, allowing a good selectivity for $\chi _{qs}/2\pi=-1.64$~MHz.

\section{The cavity's cat state preparation and Wigner tomography}

To create the WPD's coherent superposition state of $\ket{0}$ and $\ket{\alpha}$, we first drive the cavity from $\left\vert 0\right\rangle $ to the coherent state $\left\vert \alpha /2\right\rangle $ by performing a
phase-space displacement, described by the operator $D(\alpha /2)=\exp \left[
(\alpha a^{\dagger }-\alpha ^{\ast }a)/2\right] $, and transform the qubit
from $\left\vert g\right\rangle $ to the superposition state $\sin \theta
\left\vert g\right\rangle +\cos \theta \left\vert e\right\rangle $. Each of
these operations is achieved by application of a classical microwave pulse.
Then the qubit and cavity are entangled by the conditional $\pi $-phase
shift gate $U$. After a subsequent Hadamard transformation, detection of the
qubit in $\left\vert g\right\rangle $ projects the cavity to the
superposition state $\cos \theta \left\vert -\alpha /2\right\rangle +\sin
\theta \left\vert \alpha /2\right\rangle $, which is then transformed by the
displacement $D(\alpha /2)$ to the state $\left\vert \phi \right\rangle $ in
Eq.~3 in the main text. We note that the superposition state of the WPD can also be realized deterministically as described in Ref.~\onlinecite{Vlastakis} but with a slightly lower fidelity. Therefore we only present data based on a measurement-induced superposition state of the WPD. Figure~\ref{fig:figS4} shows the measured Wigner tomography of the WPD prepared in $\ket{\phi}=\mathnormal{\cos\theta}\ket{0}+\mathnormal{\sin\theta}\ket{\alpha}$ (Eq.~3 in the main text) with $\alpha=2\sqrt{2}$ and $\theta=\pi/4$. The fidelity $F=\left\langle \phi \right| \rho _p\left| \phi\right\rangle =0.93$, where $\rho _p$ is the density operator of the produced cavity state.

To verify the presence of the quantum coherence between the on and off states of the WPD, we perform the Wigner tomography after the Ramsey interference experiment~\cite
{Deleglise,Vlastakis,SunNature,Lutterbach1997}. In Figs.~\ref{fig:figS5}a and \ref{fig:figS5}b we present the reconstructed states of the cavity conditional on the qubit being measured in $\left\vert
g\right\rangle $\ and $\left\vert e\right\rangle $\ after H$_{2}$,
respectively, all for $\theta =\pi /4$ and $\varphi =\pi /2$. As expected,
each of the measured Wigner functions (upper row) exhibits interference
fringes, with alternate positive and negative values between $0$ and $\alpha
$ ($-\alpha $) in phase space, which are the signature of quantum coherence
between $\left\vert 0\right\rangle $ and $\left\vert \alpha \right\rangle $ (%
$\left\vert -\alpha \right\rangle $). This is also manifested
in the off-diagonal elements $\left\vert \rho _{n,0}\right\vert $ and $%
\left\vert \rho _{0,n}\right\vert $ in the density matrices (lower row) in the Hilbert space
obtained from the measured Wigner functions. These results unambiguously demonstrate that the qubit's behaviors with and without interference are observed with the same setup, where the WPD is in a quantum
superposition of its on and off states.

\section{Readout properties}
The readout properties of the qubit and cavity photon state are first characterized as shown in Fig.~\ref{fig:figS6}. Because we initialize the qubit to the ground state $\ket{g}$ and use pulses with DRAG to minimize the leakage to levels higher than the first excited state $\ket{e}$, we do not distinguish the levels higher than $\ket{e}$ in the readout. We adjust the phase difference between the JPA readout signal and the pump in such a way that $\ket{g}$ and $\ket{e}$ states can be distinguished with optimal contrast. Figure~\ref{fig:figS6}a shows the histogram of the qubit readout. The histogram is clearly bimodal and well separated. A threshold $V_{th}=0$ is chosen to digitize the readout signal to $+1$ and $-1$ for $\ket{g}$ and $\ket{e}$ states, respectively.

The purification of the qubit to $\ket{g}$ is achieved by a QND measurement of the qubit state; the subsequent measurement shows that the probability of the qubit being populated in $\ket{g}$ is as high as 0.996 after the initial purification (dashed histogram in Fig.~\ref{fig:figS6}a). Figure~\ref{fig:figS6}c shows the basic qubit readout properties with the cavity left in vacuum. The readout fidelity of the qubit state $\ket{e}$, prepared by properly pumping the selected $\ket{g}$ state (Fig.~\ref{fig:figS6}d), is 0.943. The readout errors are mainly due to the $T_1$ process during the waiting time after the initialization measurement (240~ns) and during the readout time (240~ns).

Figure~\ref{fig:figS6}e shows the readout property of the cavity parity state. Due to the decoherence of the qubit, there is a slight difference (typically a couple of percent) between the cases starting with $\ket{g}$ or $\ket{e}$. Here we neglect this small difference. The cat states with $\alpha =2\sqrt{2}$ and the vacuum state $\ket{0}$ are created through pulse sequences in Fig.~\ref{fig:figS3} with the appropriate $\theta$ and $\varphi$. The parity readout infidelity for the cat states is mainly due to the loss of effectiveness of the $\pi/2$ pulses and the qubit decoherence during the parity measurement (Fig.~\ref{fig:figS6}f).

Figure~\ref{fig:figS6}g shows the selectivity property of the Gaussian $\pi$ pulse $R_{\pi,0}^Y$ with $\sigma =100$~ns for distinguishing between the cavity states $\ket{\pm \alpha}$ and $\ket{0}$ (Fig.~\ref{fig:figS6}h). For $\alpha =2\sqrt{2}$, the difference between the readout fidelities of the cavities state $\ket{\pm \alpha}$ associated with the qubit states $\ket{g}$ and $\ket{e}$ is due to the decay of $\ket{e}$ during the long duration of this selective $\pi$ pulse. Here we note that the cavity states $\ket{\pm \alpha}$ and $\ket{0}$ are created with and without a direct displacement $D(\pm \alpha)$. The fidelities will be a few percent lower if we create $\ket{\pm \alpha}$ and $\ket{0}$ states using the sequence in Fig.~\ref{fig:figS3} with $\theta=\pi/2$ and 0, respectively.

\section{Quantum erasure}
\begin{figure}
\includegraphics{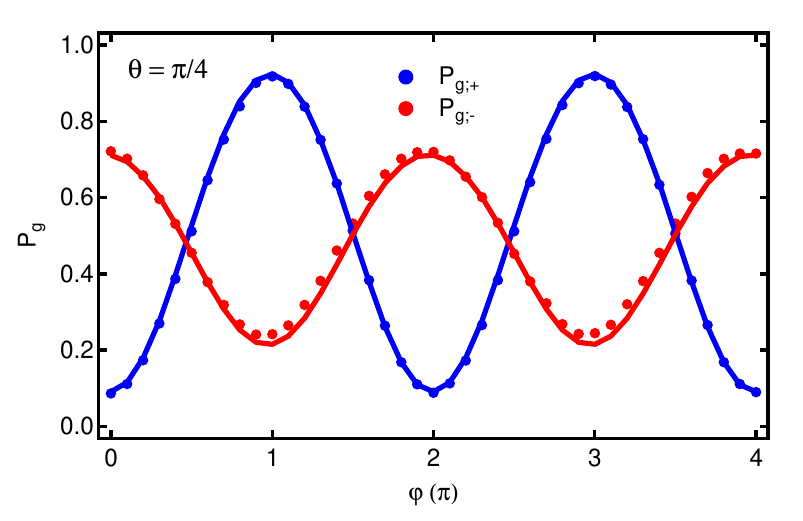}
\caption{\textbf{Conditional Ramsey interference patterns based on the cavity parity measurement.} P$_{g;+}$\ and P$_{g;-}$ represent the probabilities as a function of $\varphi$ for detecting the qubit in the state $\left| g\right\rangle $\ in the Ramsey interference experiment conditional upon the detection of even and odd parities of the cavity field, respectively. Due to the erasure of the which-path information, the interferences reappear. Dots are experimental data with the standard deviation less than 0.01 and not shown, in good agreement with the numerical simulations (solid lines) based on the measured device parameters.}
\label{fig:figS7}
\end{figure}

In our quantum delayed-choice experiment, the which-path information associated with the qubit's particle-like behavior is encoded in the phase of the cavity field, which can be erased by performing the parity measurement on the cavity even after the qubit has been detected~\cite{Scully1982,Scully1991,Gerry1996,Herzog1995,Kim2000,Ma2013}. To see this clearly, we rewrite the output state of the total system after H$_2$ in the Ramsey interference measurement as 
\begin{equation}
\ket{\psi} = \ket{\phi _{+}}\left(\cos\theta \ket{0}+\frac{\sin\theta \ket{\Phi _{+}}}{\sqrt{2}}\right)+\ket{\phi _{-}}\frac{\sin\theta \ket{\Phi_{-}}}{\sqrt{2}},
\end{equation}
where $\left| \phi _{\pm }\right\rangle =\frac 12[(\left| g\right\rangle
+\left| e\right\rangle )\pm e^{i\varphi }(\left| e\right\rangle -\left|
g\right\rangle )]$ and $\left| \Phi _{\pm }\right\rangle =\frac 1{\sqrt{2}%
}\left( \left| \alpha \right\rangle \pm \left| -\alpha \right\rangle \right) 
$. This implies that the probability for finding the qubit being in the
state $\left| g\right\rangle $ conditional upon the parity of the WPD being
even or odd is $P_{g;\pm }=\frac 12\left( 1\mp \cos \varphi \right) $, which
exhibits interference fringes with a unit contrast. The results have a
simple explanation. Each of the two coherent states $\left| \alpha \right\rangle $ and $\left| -\alpha \right\rangle $ consists of both even- and odd-parity components $\left| \Phi _{+}\right\rangle $ and $\left| \Phi_{-}\right\rangle $ with the same weighting; the which-path information is encoded in the states $\left| \alpha \right\rangle $ and $\left| -\alpha \right\rangle $, other than in $\left| \Phi _{+}\right\rangle $ and $\left|
\Phi _{-}\right\rangle $. When the cavity is detected in a definite
parity state, it is impossible to determine whether it comes from $\left|
\alpha \right\rangle $ or from $\left| -\alpha \right\rangle $, erasing the which-path information. Fig.~\ref{fig:figS7} displays the measured conditional probabilities $P_{g;+}$\ and $P_{g;-}$\ as functions of $\varphi $\ for $\theta =\pi /4$, which are defined as the probabilities for detecting the qubit in the state $\left| g\right\rangle $\ after H$_2$, conditional upon the detection of even\ and odd parities\ of the WPD, respectively. The measurement sequence is identical to Fig.~\ref{fig:figS3}a except for removing the ``WPD on/off selection" (M5). As expected, $P_{g;+}$\ and $P_{g;-}$\ exhibits complementary interference patterns, that is, the maxima in $P_{g;+}$ correspond to the minima in $P_{g;-}$. The fringe visibilities associated with these two interference patterns are 0.82 and 0.50, respectively. The reduction of the fringe contrast, compared to the ideal case, can be explained as follows. Due to the finite bandwidth of the pulses used to perform the qubit $\pi/2$ rotations in the frequency domain, when the photon number of the cavity increases, the fidelity of these rotations decreases, resulting in the reduction of the parity measurement fidelity. The vacuum state constitutes the main contribution to the even parity component, so that the Hadamard gates operate more effectively for the even parity than for the odd parity, which accounts for the result that the fringe contrast of $P_{g;+}$ is higher than that of $P_{g;-}$.

\section{Analysis of errors in marking the which-path information}
As has been shown in the main text, the procedure for marking the which-path information consists of a displacement operation $D(-\alpha)$, a conditional qubit $\pi$\ rotation $R_{\pi,0}^{Y}$, and a qubit state readout (Fig.~\ref{fig:figS3}b). Due to the dispersive qubit-cavity coupling, the cavity frequency depends on the qubit state, so that the cavity coherent states associated with the qubit state $\left|e\right\rangle $ are asynchronous with those associated with $\left|g\right\rangle $. To make the coherent state discrimination independent of the qubit state, the displacement operation is delayed in such a way that the coherent states associated with different qubit states re-synchronize when the displacement pulse is applied. To meet the required long delay time, we choose $4\sigma=680$~ns for $R_{\pi,0}^{Y}$, which also gives a better selectivity between $\ket{0}$ and $\ket{\alpha}$. The waiting time after the measurement is properly adjusted by checking the alignment of centers of the coherent states in the phase space associated with $\ket{g}$ and $\ket{e}$ states. 

Without considerations of experimental imperfections, neither conditional qubit $\pi$ rotation $R_{\pi,0}^{Y}$ can cause the qubit state to flip when the cavity is in the coherent state $\left| -\alpha \right\rangle $\ after the Ramsey interference experiment. Therefore, the conditional probability $P_{g;O,-\alpha}$ in Fig.~4c of the main text is
\begin{equation}
P_{g;O,-\alpha }=\frac{P_{g,g,g}}{P_{g,g,g}+P_{e,e,e}}, 
\end{equation}
where $P_{g,g,g}$($P_{e,e,e}$) is the probability that the qubit is measured to be in $\left| g\right\rangle $ ($\left| e\right\rangle $) state during all the processes in the last three measurements (M4-M6) in Fig.~\ref{fig:figS3}b, including observation of the Ramsey interference pattern, post-selection of the WPD's action, and distinction between $\left| \alpha \right\rangle $ and $\left| -\alpha \right\rangle $. The ideal values of $P_{g,g,g}$ and $P_{e,e,e}$ are both equal to $1/8$. However, due to the qubit energy decay, $P_{e,e,e}$ becomes smaller so that $P_{g;O,-\alpha}>1/2$. This decay (on average about 1.2~$\mu$s long) mainly occurs during the long duration for performing the conditional qubit $\pi$ rotation, the measurement time, and the delay for re-synchronizing the joint qubit-cavity state. The measured average value of $P_{g;O,-\alpha}$ is 0.56, in good agreement with the expected one taking into account the qubit decay: $P_{g,g,g}/(P_{g,g,g}+P_{e,e,e}) \approx 1/(1+e^{-1.2\times 2/9.5})=0.56$. The small fringe contrast mainly comes from the residual vacuum which can be further excluded by performing additional operations $D(\alpha ) $ and $R_{\pi ,0}^{Y}$ and then measuring the qubit state. 

On the other hand, when the cavity is measured to be in the coherent state $\left| \alpha \right\rangle$ after the coherent state discrimination, the corresponding conditional probability $P_{g;O,\alpha}$ shows almost no oscillation (the fringe contrast is less than 0.01). The reason is that $\ket{\alpha}$ state is nearly perfectly distinguished from $\ket{-\alpha}$ and $\ket{0}$ in this process. As a result, $P_{g;O,\alpha} $ is given by
\begin{equation}
P_{g;O,\alpha}=\frac{P_{g,g,e}}{P_{g,g,e}+P_{e,e,g}}, 
\end{equation}
where $P_{g,g,e}$ ($P_{e,e,g}$) denotes the probability that the qubit is measured to be in $\left| g\right\rangle $ ($\left| e\right\rangle $) state during the observation of the Ramsey interference pattern and the subsequent post-selection of the WPD's action, and in $\left| e\right\rangle$ ($\left| g\right\rangle $) during the procedure for distinguishing between $\left| \alpha \right\rangle $ and $\left| -\alpha \right\rangle$ (M4-M6 in Fig.~\ref{fig:figS3}b). The measured average value of $P_{g;O,\alpha}=0.52$, with the small deviation from 0.5 mainly coming from the qubit decay during the post-selection of the WPD's action ($\approx 1/(1+e^{-1/9.5})=0.53$). We note that the cavity state $\ket{\alpha}$ is transformed to $\ket{0}$ during the coherent state discrimination, so that the readout fidelities of this cavity state associated with the qubit states $\ket{e}$ and $\ket{g}$ are almost the same.

\bibliography{bibliography}

\end{document}